\documentclass{aa}  

\usepackage{tabularx}
\usepackage{graphicx}
\usepackage{xcolor}
\usepackage{txfonts}
\usepackage[hyperfootnotes=true]{hyperref}
\usepackage{tablefootnote}
\usepackage{lipsum}
\usepackage{subcaption}         
                               
\usepackage{lscape}             
                                
\usepackage{placeins}            
                                
\titlerunning{Impact cratering in the Uranian satellite system}

\begin{document}

   \title{Impact cratering in the Uranian satellite system: \\ 
  assessing its main impactor populations}

   \author{N. L. Rossignoli\inst{1}\corrauth{nrossignoli@fcaglp.unlp.edu.ar}   
        \and R. P. Di Sisto\inst{1,2}\email{romina@fcaglp.unlp.edu.ar}
        }

   \institute{Instituto de Astrofísica de La Plata, Facultad de Ciencias Astronómicas y Geofísicas, Universidad Nacional de La Plata, CCT La Plata - CONICET - UNLP, Paseo del Bosque S/N (1900), La Plata, Argentina \and Facultad de Ciencias Astronómicas y Geofísicas, Universidad Nacional de La Plata, Paseo del Bosque S/N (1900), La Plata, Argentina}

   \date{Received May 29, 2026; Accepted Nov 23, 2026}

\abstract
  {The Uranian satellite system presents a unique scenario to test impact cratering models for the outer Solar System. Its extreme axial tilt provides insights into the system's early history, possibly shaped by a giant impact. This event may have generated a planetocentric impactor population that played a significant role in shaping the observed crater distributions of the satellites, in addition to contributions from heliocentric sources.}
  {This work aims to explore and assess, using theoretical models, the contributions of both heliocentric and planetocentric impactor populations in the cratering history of the regular Uranian satellites.}
 {A theoretical model previously developed for the Saturnian satellite system was adapted and applied to compute the crater production by heliocentric impactors on the regular Uranian satellites, considering the uncertainty in their size-frequency distribution (SFD) and the current configuration of the Solar System (thus excluding the early post-instability bombardment). In addition, planetocentric impactor populations were modeled based on a debris disk generated by a giant impact, with their SFDs calibrated using Oberon's observed crater distribution. The resulting theoretical crater distributions were compared  with their observational counterparts for each mid-sized Uranian satellite.}
 {Crater distributions produced by our modeled heliocentric impactor population are found to be insufficient to explain the highly cratered surfaces of Umbriel, Titania, and Oberon. In contrast, the results obtained with the planetocentric impactor population model are able to reproduce the observed distributions more accurately, especially for highly cratered terrains on Oberon and Umbriel. Across certain diameter ranges, overestimation of the observed craters may be related to resurfacing processes.}
 {The dominant impactor population in the Uranian satellite system is found to be likely comprised by planetocentric impactors formed from a giant impact, rather than by the modeled heliocentric impactors, with resurfacing and geological activity further shaping observed crater distributions on each satellite.}

 
   \maketitle
   \nolinenumbers

\section{Introduction}
\label{intro}

The Uranian satellite system presents unique aspects to explore and test impact cratering models in the outer Solar System. The planet, its ring system and the regular satellites all present an extreme axial tilt that has long been linked to a giant impact event in an early stage of the planet's evolution \citep[e.g.,][]{Safronov1966}. The Voyager 2 mission's flyby of this system, in January 1986, provided the only images of the surfaces of the satellites available to date. Only the mid-sized satellites Miranda, Ariel, Umbriel, Titania and Oberon were resolved, together with the inner satellite Puck. Due to the orientation of the Uranian system at that time, only the southern hemispheres of the moons were imaged, with varying resolutions \citep[e.g.,][]{Strom1987,Kirchoff2022}.
The impact crater distributions of the Uranian satellites obtained from these images were not unique, but varied among different authors \citep[e.g.,][]{Plescia1987UTO,Strom1987}. 
Nevertheless, the cratering record did show at least two distinct crater populations of different ages \citep{Strom1987}. While Oberon and Umbriel exhibit heavily cratered surfaces, Titania and Ariel appear to have undergone endogenic resurfacing \citep{Plescia1987UTO,Plescia1987Ar}. In addition, Miranda displays a very complex geology, with its surface divided into heavily cratered terrain and resurfaced regions associated with large tectonic features. While heliocentric impactors such as Centaur objects are known to be pervasive across the outer Solar System, Uranus' axial tilt raises the possibility of a substantial planetocentric debris population originating from a large impact event.

In this work, we present a detailed study of crater production on the Uranian regular satellites using updated impact cratering models. On the one hand, we use previous studies on cratering rates in the Saturnian system \citep{DiSisto2011,DiSisto2013,Rossignoli2019,Rossignoli2022} to model the contribution of heliocentric impactors to the crater distributions on the regular Uranian satellites. On the other hand, we study the contribution of planetocentric impactors following recent studies that explore the formation of the mid-sized Uranian satellites from a debris disk generated from the possible giant impact that tilted the planet \citep{Kegerreis2018,Ida2020,Woo2022}. In this scenario, the planetocentric impactors emerge from the final stage of the regular satellites’ formation, during which the debris disk from which they accreted was still dynamically excited and contained objects that could become potential satellite impactors.

In Sect.~\ref{Sec2} we describe the Uranian satellite system, its possible origin linked to a giant impact, and the current knowledge on observed crater distributions on the mid-sized satellites. In Sect.~\ref{SecMethod} we describe the heliocentric and planetocentric impact cratering models. In Sect.~\ref{secresul} we describe the results obtained with our impact cratering model and the comparisons with the observed distributions. Finally, in Sect.~\ref{SecConclusions} we discuss the role of different impactor populations in this system and the constraints on resurfacing processes for the regular Uranian satellites.

\section{The Uranian satellite system} 
\label{Sec2}

\subsection{The Uranian ring system and inner satellites}

Uranus' ring system consists of 13 distinct rings, including ten narrow rings (6, 5, 4, $\alpha$, $\beta$, $\eta$, $\gamma$, $\delta$, $\lambda$, $\epsilon$) and three diffuse rings ($\zeta$, $\nu$, $\mu$). Most of the narrow rings were discovered through stellar occultations in 1977, while others were later identified in images from the Voyager 2 mission and the Hubble Space Telescope.

Among them, the $\epsilon$ ring is the most prominent, showing a noticeable eccentricity and significant radial width variations between periapsis and apoapsis. The diffuse $\nu$ and $\mu$ rings extend significantly farther outward, with the $\mu$ ring reaching distances close to 100\,000 km.

The 14 small inner satellites of Uranus were mostly discovered during the Voyager 2 flyby in 1986, with additional detections made later using reanalyzed data and Hubble observations. Recently, a small satellite designated Uranus XXVIII was discovered by \citet{ElMoutamid2025} using the James Webb Space Telescope. These satellites have radii between a few kilometers and about 80 km and orbit within a relatively compact region close to the rings.

Next, we describe these satellites ordered by increasing distance from Uranus. Table~\ref{urainnprop} lists their general properties.

\begin{table}[h!]
\caption{General properties of the inner Uranian satellites} 
\label{urainnprop}   
\centering                      
\begin{tabular}{lcccc}    
\hline\hline\noalign{\smallskip}   
Satellite & $R_\text{m}$ & $a$  & $e$ & $i$  \\
& [km] & [km] & & [deg] \\
\hline\noalign{\smallskip}
  Cordelia & 21  & 49\,755 & 0.000 & 0.2 \\ 
  Ophelia   & 23 & 53\,765 & 0.011 & 0.1 \\ 
  Uranus XXVIII & 4-5 & 57\,844 & 0.039 & 4 \\
  Bianca   & 27   &  59\,170 & 0.006 & 0.1 \\
  Cressida  & 41   &  61\,770 & 0.004 & 0.1 \\
  Desdemona & 35   &  62\,663 & 0.007 & 0.1 \\
  Juliet  & 53   &  64\,362 & 0.006 & 0.0 \\
  Portia   & 70   & 66\,101 & 0.004 & 0.0 \\
  Rosalind & 36   &   69\,930 & 0.003 & 0.0 \\
  Cupid   & 9   &  74\,396 & 0.007 & 0.1 \\ 
  Belinda  & 45   &  75\,258 & 0.002 & 0.0 \\
  Perdita  & 13   &  76\,418 & 0.005 & 0.0 \\
  Puck     & 81   &  86\,007 & 0.009 & 0.3 \\
  Mab      & 6-12    &  97\,737 & 0.006 & 0.1 \\
\hline                         
\end{tabular}
\tablefoot{Mean radius $R_\text{m}$ from \citet{French2024}, except for Uranus XXVIII from \citet{ElMoutamid2025}, and Cupid, Perdita and Mab from \citet{Showalter2006}; semimajor axis $a$, eccentricity $e$, and inclination $i$ with respect to the planet's equatorial plane from the JPL planetary satellite mean elements file\tablefootnote{ \url{https://ssd.jpl.nasa.gov/sats/elem/sep.html}}.}
\end{table}

\subsubsection{Ring shepherd satellites}

Cordelia ($R_{\text{m}} \!\sim\!$ 21~km) and Ophelia ($R_{\text{m}} \!\sim\!$ 23~km) orbit just inside and outside the $\epsilon$ ring, respectively. These moons act as shepherd satellites, confining the ring through gravitational torques that maintain its sharp edges and discrete resonances \citep[e.g.,][]{Showalter2020}.

\subsubsection{Uranus XXVIII}
The newly discovered satellite, initially designated S/2025 U1, was detected in images obtained with the Near Infrared Camera onboard NASA's James Webb Space Telescope on February 2, 2025 \citep{ElMoutamid2025}. With an estimated radius between 4-5 km, this small satellite orbits between Ophelia and Bianca \citep{ElMoutamid2025}.  

\subsubsection{The Portia group}

A set of nine satellites, from Bianca to Perdita, forms a closely packed dynamical group known as the Portia group. These moons orbit within a narrow radial region (see Table~\ref{urainnprop}), with separations of less than 4500 km between neighboring objects \citep{Smith1986,Showalter2020}.
Their proximity leads to a complex network of mean-motion resonances, making the system dynamically active and potentially unstable over long timescales \citep{DuncanLi1997,French2012}. Numerical simulations suggest that some pairs, such as Cupid-Belinda and Cressida-Desdemona, may eventually cross orbits and collide on timescales ranging from $10^3$ to $10^7$ years \citep{French2012}.

\subsubsection{Puck and Mab}
\label{Pmabsec}
Puck ($R_{\text{m}} \!\sim\!$ 81~km) is the largest of the inner satellites and exhibits a heavily cratered surface \citep{Smith1986}, with its largest detected crater reaching a diameter of $\sim\!$ 47 km \citep[Bogle crater,][]{Smith1986}. In addition, its surface may be partially covered by infall of particles from the neighboring $\mu$ ring \citep{Showalter2020,Cartwright2021}.

Mab, on the other hand, is one of the smallest known regular Uranian satellites, with an estimated radius between 6 and 12 km \citep{Showalter2006}. It orbits within the $\mu$ ring, which may be composed of material ejected from its surface. Recent observations suggest that Mab has a blue spectral signature similar to Miranda, supporting the smaller size estimate of a 6 km radius \citep{Molter2023}.

\subsection{Mid-sized satellites}

The discovery of Uranus’ major satellites began in 1787 with Titania and Oberon, identified by William Herschel shortly after the discovery of the planet itself. Ariel and Umbriel were later discovered by William Lassell in 1851, while Miranda, the smallest and innermost of the five, was identified by Gerard Kuiper in 1948. These satellites have mean radii between approximately 235 and 789 km, with the largest and most massive objects orbiting farther from Uranus. Collectively, they account for about 0.01\% of the planet’s mass. Their compositions are thought to consist of roughly equal proportions of ice and rock, except for Miranda, which is believed to be predominantly icy \citep{Jacobson2022}.

Next, we describe these satellites ordered by increasing distance from Uranus. Table~\ref{uramedprop} lists their general properties.

\begin{figure}
	\centerline{\scalebox{0.43}
		{\includegraphics{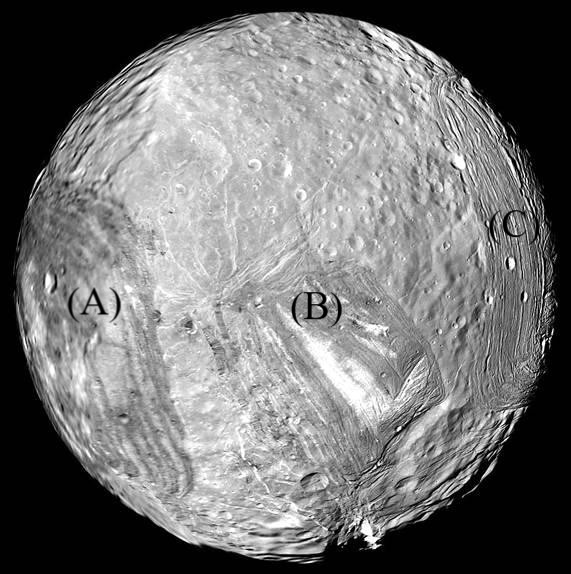}}}
			\caption{Miranda as seen by Voyager 2 on January 24, 1986. The three coronae are indicated in the image: Arden (A), Inverness (B) and Elsinore (C). Base image credit: PIA18185, NASA/JPL-Caltech.}
		\label{MirandaNASA}
	\end{figure}

Miranda is the smallest and innermost of the mid-sized, ``Classical'' satellites, and also the least dense (see Table~\ref{uramedprop}). Voyager 2 approached the satellite within 29\,000 km, capturing the highest-resolution images of any Uranian moon and thus revealing a geologically complex surface. Among the most striking features are three polygonal coronae: Arden, Inverness, and Elsinore (Fig.~\ref{MirandaNASA}) that are indicative of tectonic and/or cryovolcanic processes \citep{Beddingfield2020}. These structures are unique to Miranda and suggest significant internal activity \citep{Schenk2020}. In addition, the presence of heavily cratered regions indicates that terrain ages vary widely across different areas \citep{Kirchoff2022}. Similarly, Ariel shows signs of widespread surface renewal \citep{Schenk2020}. Voyager 2 images revealed extensive chasmata, fracture systems, and grooved terrain associated with tectonic and/or
cryovolcanic activity. Moreover, the satellite’s craters exhibit varying degrees of degradation such as in the case of Domovoy and Yangoor, two similarly sized craters that differ both in depth and morphology, with the latter showing signs of viscous relaxation \citep{Schenk2020}.

Umbriel, although similar in size to Ariel, displays a darker surface with low albedo \citep{Smith1986}. Voyager 2’s observations revealed polygonal fracture basins and a high density of impact craters, suggesting an older surface \citep{Schenk2020}. Titania, the largest Uranian moon, showed large tectonic fractures and extensive cratering \citep{Schenk2020,Nathan2024}. Its largest identified crater, Gertrude, is about 326 km and presents a relatively shallow depth ($\sim$1-3 km), suggesting that viscous relaxation (or other flattening/infill process) has modified this crater \citep{Nathan2024}. In addition, the prominent Messina Chasmata canyon system spans over 1500 km and may have formed due to interior freezing, hinting at a former subsurface ocean \citep{Cartwright2021}. Oberon has a similarly cratered surface. Its largest craters, Hamlet and Othello, are associated with bright ejecta patterns, and an 11 km-high elevation feature may be the central peak of a complex crater or the result of tectonic or volcanic processes \citep{Schenk2020}.

\begin{table}[h!]
\caption{General properties of the mid-sized Uranian satellites}      
\label{uramedprop}    
\flushleft      
\noindent 
\begin{tabular}{lcccccc}      
\hline\hline\noalign{\smallskip}   
Satellite & $R_\text{m}$ & $M$ & $\rho$ & $a$ & $e$ & $i$ \\
& {\scriptsize [km]} & {\scriptsize [10$^{20}$ kg]} & {\scriptsize [g\,cm$^{-3}$]} & {\scriptsize [km]} & & {\scriptsize [deg]} \\       
\hline\noalign{\smallskip}         
Miranda & 235.8 & 0.66 & 1.18 & 129\,846 & 0.001 & 4.42 \\
Ariel & 578.9 & 12.9 & 1.54 & 190\,929 & 0.001 & 0.03 \\
Umbriel & 584.7 & 12.2 & 1.52 & 265\,986 & 0.004 & 0.08 \\
Titania & 788.9 & 34.2 & 1.65  & 436\,298 & 0.002 & 0.11 \\
Oberon & 761.4 & 28.8 & 1.66 & 583\,511 & 0.002 & 0.13 \\
\hline                            
\end{tabular}
\tablefoot{Mean radius $R_\text{m}$ from \citet{Archinal2018}, mass $M$ from \citet{Jacobson2025}, mean density $\rho$ from the JPL planetary satellite physical parameters file\tablefootnote{\url{https://ssd.jpl.nasa.gov/sats/phys_par/}} (with errors in the range 0.03--0.05 g\,cm$^{-3}$),
semimajor axis $a$, eccentricity $e$, and inclination $i$ with respect to the planet's equatorial plane from \citet{Jacobson2025}.}
\end{table}

\subsubsection*{Possible origins}

Regarding the formation of these satellites, several studies \citep[e.g.,][]{Kegerreis2018, Ishizawa2019, Ida2020, Woo2022} have investigated their connection to the origin of the system's obliquity. In this scenario, a giant impact between a proto-Uranus and an impactor with a mass of 2–3 Earth masses could lead to the current tilt of the planet's spin axis and simultaneously generate a debris disk from which the regular satellites could accrete in the planet's equatorial plane. However, \citet{Morbidelli2012} showed that the prograde orbits of the regular satellites require Uranus to have had a non-negligible initial obliquity, comparable to that of Neptune, before its final tilting to the present value.

\citet{Safronov1966} was among the first to propose that Uranus' current obliquity could be linked to a giant collision. Later, SPH simulations by \citet{Kegerreis2018} showed that such an impact could eject material beyond Uranus’ corotation radius (3.30 $\text{R}_\text{U}$, where R\textsubscript{U} is the radius of Uranus), forming a planetocentric debris disk. Based on these results, \citet{Ishizawa2019} modeled satellite accretion via N-body simulations but found that the resulting systems failed to reproduce the observed architecture, suggesting the need for post-impact disk evolution.
Subsequently, \citet{Ida2020} modeled the coupled viscous spreading and cooling of an initially vaporized impact-generated disk, showing that ice condensation in the outer regions leads to a positive surface density gradient and reproduces the observed mass-orbit configuration.

 \citet{Reinhardt2020} performed SPH simulations that showed that giant impacts can produce extended debris disks, with a significant fraction of the material located outside the Roche radius shortly after the collision. This orbiting material may contribute to satellite formation, suggesting that Uranian satellites could form approximately in situ following the viscous spreading and evolution of the disk. In order to further investigate this scenario, \citet{Woo2022} combined hydrodynamic simulations to model an oblique impact between a proto-Uranus and a rocky impactor of 2-3 Earth masses, a semi-analytical model to study the post-impact disk evolution, and numerical simulations to form satellites from the disk. Based on this approach, they successfully reproduced the current configuration of the Uranian mid-sized satellite system, including the mass distribution (increasing mass with distance, see Table \ref{uramedprop}) and the general composition of equal parts of ice and rock for the four outermost satellites and a predominantly icy composition for Miranda. More recently, \citet{Murashima2026} showed that, although the mass and size of the impact-generated disk are relatively robust to modeling choices, its rock fraction strongly depends on the adopted equation of state (EOS), highlighting the model dependence of the predicted disk composition.

\subsection{Observed crater distributions on the Uranian satellites}
The Uranian system was only imaged up-close by Voyager 2 in January 1986. The spacecraft flew by the system collecting images that revealed the presence of 10 new satellites and 2 new planetary rings, among other discoveries. Due to the orientation of the Uranian system at that time, only the southern hemispheres of Puck and the mid-sized satellites were resolved, with resolutions varying between 0.26 km/pixel on Miranda to 6.6 km/pixel on Oberon \citep{Smith1986, Strom1987, Cartwright2021, Kirchoff2022}.

During the Voyager era, several studies analyzed these images and provided the first characterizations of the surfaces of the classical satellites. Although the derived crater distributions differed among authors \citep[e.g.,][]{Plescia1987UTO,Strom1987}, these early analyses revealed at least two distinct crater populations of different ages \citep{Strom1987} and established the basis for subsequent investigations. More recently, \citet{Kirchoff2022} reassessed the crater populations on the five classical satellites using reprocessed images from the Voyager 2 Imaging Science Subsystem library, which covered between 35\% and 45\%  of the satellites' surfaces. Due to the varying image resolutions, they obtained complete crater counts down to diameters of $1.3-1.7$ km for Miranda, 6.3 km for Ariel, 25.4 km for Umbriel, $15.5-17.5$ km for Titania, and 33.0 km for Oberon. 

\section{Method}
\label{SecMethod}   

Based on previous works that model crater distributions on the Saturnian satellites \citep{DiSisto2011,DiSisto2013,Rossignoli2019,Rossignoli2022}, we calculate the theoretical production of craters on the regular Uranian satellites considering initially only heliocentric impactors. In this first approach, the impactor population is modeled with Centaurs that evolve from the Scattered Disk (SD) in the Transneptunian Region (TNR) into inner regions of the Solar System. In addition, in the present work we model different planetocentric impactor populations to assess their role in the production of craters on the five mid-sized Uranian satellites. These planetocentric impactor populations are built on the basis of recent models that link the formation of the regular satellites to the possible giant impact that tilted the planet’s spin axis.

In this section we describe the theoretical cratering method developed to consider the contribution of both heliocentric and planetocentric impactors to the crater distributions on the mid-sized satellites.

\subsection{Heliocentric impactor population model}
\label{heliomodel}
We model the heliocentric impactor population considering Centaurs originating in the SD, which, due to perturbations exerted mainly by Neptune, migrate inward into the planetary region, exhibiting a transient dynamical nature. Based on the study by \citet{DiSisto2020} and previous works \citep[e.g.,][]{Levison1997, DiSisto2007}, the numerical simulations show that within the TNR, the subpopulation with the highest probability of having encounters with Neptune and therefore of evolving toward the planetary region of the Solar System is the scattered disk objects (SDOs). Once the SDOs migrate into the giant planetary region, they become Centaurs, which constitute the main heliocentric impactor population considered in this work. 

To compute the impact rates of Centaurs with the regular satellites of Uranus, we use the results of the simulation presented in \citet{DiSisto2020}. In this work, the authors performed a numerical integration of an intrinsic model of the SD, considering the current configuration of the Solar System and following the evolution of the particles during 4.5 Gyr. Thus, the simulation does not account for the early dynamical instability of the giant planets and the associated substantial depletion of the primordial trans-Neptunian disk. Instead, it follows the subsequent evolution of SDOs once the stabilization phase had taken place.  

As in previous studies \citep[e.g.,][]{Rossignoli2022}, we consider that the cumulative size-frequency distribution (SFD) of the impactors is given by a power law with a break at diameters $d=100$ km:

\begin{xalignat}{4}
\label{distroCSD}
        N(>d) &= C_0 \,\bigg(\frac{1 \text{ km}}{d}\bigg)^{s_2 - 1} &&\text{for} && d \leq 100~\text{km},  \nonumber \\
        N(>d) &= \text{3.5} \times 10^{5} \, \bigg(\frac{100 \text{ km}}{d}\bigg)^{s_1-1} &&\text{for} && d > 100~\text{km},
\end{xalignat} 

\noindent where $C_0 = \text{3.5} \times 10^{5} \times 100^{s_2-1}$ by continuity at $d$ $=$ 100 km, $s_1 = 4.7$ is the differential size distribution index for the brighter objects, and $s_2$ is the differential size distribution index for the smaller impactors that is modeled with two limiting values: $s_2 = 3$ and $ 3.5$. The shape of this SFD is based on multiple studies \citep{Bernstein2004, GilHutton2009, Fraser2009, Fuentes2008, Fuentes2009}. Following the observations by the Outer Solar System Origins Survey (OSSOS) \citep{Bannister18}, the study by \citet{Lawler2018} found that the size distribution of SDOs is best described by a power law with a break at diameters $d \sim 100$ km. In addition, they found limiting slopes of the absolute magnitude distribution for the fainter objects that correspond to values of $s_2 =$ 3 - 3.5 in Eq.~\ref{distroCSD}. For the larger objects, \citet{Elliot2005} obtained a differential size distribution index for SDOs of $s_1 = 4.7$. With respect to constraints on the number of SDOs, 
\citet{Parker2010b,Parker2010a} determined that the maximum total number of SDOs with diameters $d > 100$ km is $3.5 \times  10^5$ objects, which is represented by the $C_0$ value in Eq.~\ref{distroCSD}.

The results of the simulation by \citet{DiSisto2020} provide the number of encounters of SDOs with Uranus within the Hill sphere of the planet. Then, the number of collisions on each satellite is computed considering a particle-in-a-box approximation with:

\begin{equation}
N_c(> d) =  \frac{v_i\,R_G^{2}}{v\,(R_{\text{H}})\,R_{\text{H}}^{2}} 39.491\,N(> d),
\label{Nc10}
\end{equation}

\noindent where $v_i$ is the relative collision velocity on each satellite, v($R_{\text{H}}$) is the Centaurs' mean relative encounter velocity when they enter Uranus' Hill sphere (of radius $R_{\text{H}}$), and $R_G$ is the satellite's collision radius given by $R_G=R_\text{m}\, (1+(v_{esc}/v(R_{\text{H}}))^2)^{1/2}$, where $v_{esc}$ is the satellite's escape velocity. This expression accounts for the gravitational focusing exerted by the satellite itself, which increases its effective collisional cross section. The factor 39.491 in Eq. \ref{Nc10} represents the average number of passages through Uranus' Hill sphere per initial SDO in the numerical simulation \citep{DiSisto2020}. It is obtained by dividing the total number of encounters within Uranus' Hill sphere by the total number of simulated particles, including those that never encounter Uranus.

For the final step, we take into account that the computation of the impact crater size in terms of the collision parameters typically considers three stages: contact and compression, excavation, and collapse. Initially, the impactor compresses the target, generating shock waves and ejecting material. This is followed by the excavation stage, where a hemispherical shock wave forms a transient crater until the energy is insufficient to overcome the target’s resistance, either due to cohesion or gravity. Finally, gravity causes the transient crater to collapse, shaping the final crater. This last stage is especially significant in large craters and is responsible for the formation of features like central peaks. The final crater size thus depends on both the target’s gravity and material properties, requiring models to account for both excavation and collapse.

Thus, we use the crater scaling law presented in \citet{Holsapple2007} to compute the size of the transient crater $D_{\text{t}}$ produced by each impactor of diameter $d$:

\begin{equation}
D_{\text{t}} = K_{1}\left[\left(\frac{gd}{2v^{2}_{\text{in}}}\right)\left(\frac{\rho_{\text{t}}}{\rho_{\text{i}}}\right)^{\frac{2\nu}{\mu}}
+ K_{2}\left(\frac{Y}{\rho_{\text{t}}v^{2}_{\text{in}}}\right)^{\frac{2+\mu}{2}}\left(\frac{\rho_{\text{t}}}{\rho_{\text{i}}}\right)^{\frac{\nu(2+\mu)}{\mu}}\right]^{-\frac{\mu}{2+\mu}} d,
\label{HH07}
\end{equation}

\noindent where $g$ and ${\rho_{\text{t}}}$ are the target's gravity and density, respectively, and $v_{\text{in}}$ is the normal component of the impact velocity $v_{\text{i}}$, given by $v_{\text{in}}=v_{\text{i}}\sin\theta$, where we adopt $\theta=45^\circ$ measured from the horizon as the most probable impact angle. For the scaling law parameters, we consider values for icy satellites, specifically $\mu$=0.38, $\nu$=0.397 \citep[see, e.g.,][]{Kraus2011,DiSisto2013}, $K_1$=1.93, and $K_2$=1 \citep[see][]{Holsapple2022}. For the strength parameter $Y$, we consider $Y$=1 MPa for the mid-sized satellites \citep[generally used for larger icy satellites as in][]{Rossignoli2022} and $Y$=15 kPa for the smaller inner satellites \citep[see][]{Leinhardt2009,DiSisto2013,Rossignoli2019}. For the impactor density we use ${\rho_{\text{i}}}=1$ g\,cm$^{-3}$. 

With Eq. \ref{HH07} one can determine from the dominant term if, during the impact crater formation, the crater growth is limited by the target's gravity (first term) or its strength (second term). The transition crater diameter between the strength and gravity regimes $\text{D}_\text{l}$ for each satellite is listed in Table ~\ref{uramedresh}.

Finally, once the diameter of the transient crater $D_{\text{t}}$ has been obtained with Eq.~\ref{HH07}, it is necessary to consider the changes in the crater's morphology and size that occur during the final collapse stage described previously \citep{Holsapple1993}.
The transition from a simple to a complex morphology occurs at a distinct crater size given by \citet{Kraus2011}: 
\begin{equation}
D^{*} = \frac{g_{gan}}{g_T}\,2\,R^{*}_{gan}, 
\end{equation}
where  $g_{gan}=1.43$ m\,s$^{-2}$ is the surface gravity of Ganymede and $R^{*}_{gan}=1 $ km is the radius of the transition crater between simple and complex craters for Ganymede \citep{Schenk2002}. Thus, following the relation presented in \citet{Kraus2011}, the final crater diameter can be obtained with:

\begin{xalignat}{4}
 D &= (1.3\, k)D_t  &&\text{for} && D_t \leq D^*/1.3\,k,  \nonumber \\
 D &= D_t\,(1.3\,k)^{1/(1-\eta)} \left(\frac{D_t}{D^{*}}\right)^{\eta/(1-\eta)}   &&\text{for} &&  D_t > D^*/1.3\,k,
\label{dcrater}
\end{xalignat} 
where $k=1.19$ and $\eta=0.04$ \citep{Kraus2011}.

\subsection{Disruption of satellites}
\label{Secdisrup}

Given that several of the inner Uranian satellites have small sizes, we study the conditions under which a catastrophic impact could lead to their disruption. Following the model presented in previous works \citep[e.g.,][]{Rossignoli2019}, we evaluate catastrophic dispersive collisions, defined as those in which the target is fragmented and the resulting debris attains velocities exceeding the escape velocity \citep{Benz1999}.

The critical specific energy for a catastrophic dispersive event, $Q^{*}_{\text{D}}$, is defined as the specific kinetic energy required to disperse the target into a distribution of fragments such that the largest remnant retains half of the original mass. Recently, \citet{Bottke2023} investigated the collisional evolution of the destabilized primordial Kuiper Belt population, using the Boulder code \citep{Morbidelli2009} and calibrating fragmentation laws against observed size-frequency distributions of Jupiter Trojans and crater populations on the giant planet satellites. In this framework, the disruption threshold is given by \citep{Bottke2020}:

\begin{equation}
Q^{*}_{\text{D}}(R_{\text{pb}})=a\left(\frac{R_{\text{pb}}}{1\,\mathrm{cm}}\right)^{\alpha}+b\left(\frac{R_{\text{pb}}}{1\,\mathrm{cm}}\right)^{\beta},
\label{RBottke}
\end{equation}

\noindent where $R_{\text{pb}}$ is the target radius. The coefficients $a$ and $b$ are defined as:

\begin{equation*}
a=\frac{Q^{*}_{\text{D\,LAB}}}{R_{\text{LAB}}^{\alpha}}\,
\frac{1}{1-\frac{\alpha}{\beta}\left(\frac{R_{\text{LAB}}}{R_{\text{min}}}\right)^{\beta-\alpha}},
\qquad
b=-\frac{\alpha}{\beta}\,a\,R_{\text{min}}^{\alpha-\beta},
\end{equation*}

\noindent where $R_{\text{LAB}}=D_{\text{LAB}}/2$ and $Q^{*}_{\text{D\,LAB}}$ define the laboratory calibration point, and $R_{\text{min}}=D_{\text{min}}/2$ corresponds to the minimum of the function $Q\mbox{*}_{\text{D}}(R_{\text{min}})$. \citet{Bottke2023} calibrate this formulation using icy impact experiments and hydrodynamic simulations spanning impact velocities from 4 m\,s$^{-1}$ to 3.5 km\,s$^{-1}$ \citep{Leinhardt2009}, and adopting $D_{\text{LAB}}=10$ cm. Their best-fit parameters for the destabilized primordial Kuiper Belt population result in  $Q^{*}_{\text{D\,LAB}}=6.2\times10^{4}$ erg\,g$^{-1}$, $D_{\text{min}}=20$ m, $\alpha=-0.4$, and $\beta=1.2$.

This model yields significantly lower disruption thresholds than \citet{Benz1999}, primarily due to the inclusion of low-cohesion targets. As presented in \citet{Leinhardt2009}, material strength plays a critical role in determining catastrophic disruption outcomes, especially for porous bodies. Given that outer Solar System satellites may exhibit high porosity and low strength \citep{Leinhardt2009}, these prescriptions may be more suitable for small Uranian satellites.

In order to obtain the diameter of the impactor that produces a catastrophic dispersive impact, we consider the specific kinetic energy of a collision, given by:

\begin{equation}
Q=\frac{1}{2}\left(\frac{m_{\text{i}}}{M_{\text{tot}}}\right)v_{\text{i}}^{2},
\label{Rupa}
\end{equation}

\noindent where $m_{\text{i}}$ is the impactor mass and $M_{\text{tot}}$ is the total mass of the system. For a catastrophic dispersive event, we set $Q = Q^{*}_{\text{D}}$, yielding:

\begin{equation}
m_{\text{i}}=\frac{2Q^{*}_{\text{D}}M_{\text{tot}}}{v_{\text{i}}^{2}}.
\label{Rup6}
\end{equation}

Assuming a spherical impactor of density $\rho_{\text{i}}$, the corresponding disruption diameter is obtained by:

\begin{equation}
d_{\text{rup}}=2\left(\frac{6Q^{*}_{\text{D}}M_{\text{tot}}}{4\pi\rho_{\text{i}}v_{\text{i}}^{2}}\right)^{1/3}.
\label{Rup77}
\end{equation}

\subsection{Satellite age}
\label{Secedad}

Having established the conditions for catastrophic dispersive impacts, we find that some of the smaller satellites are disrupted over the age of the Solar System. Therefore, these objects cannot be primordial and must have formed or re-formed after their most recent catastrophic dispersive event. Thus, in order to estimate the time at which the last catastrophic dispersive collision occurred, we follow a previously developed method \citep{DiSisto2016,Rossignoli2019}.

\begin{figure}[h]
{\includegraphics[width=0.99\linewidth]{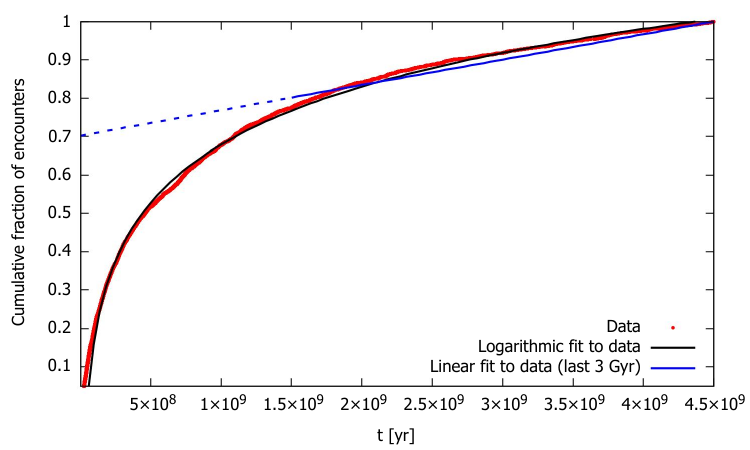}}
\caption{Cumulative fraction of encounters of Centaurs with Uranus as a function of time (red dots), together with a logarithmic fit to data (black solid line) and a linear fit to data for the last 3 Gyr, shown as a blue solid line over its intended fitting range (t > 1.5 Gyr) and as a blue dashed line outside this range.}
\label{Encur}
\end{figure}

Based on the results of the dynamical evolution of the impactors presented in \citet{DiSisto2020}, we can fit a temporal function to the cratering process, considering that the cumulative number of craters produced by Centaurs on the Uranian satellites is proportional to the cumulative number of encounters between Centaurs and Uranus. Therefore, the dependence of cratering with time is the same as that of the encounters of Centaurs with Uranus. As in previous works \citep{DiSisto2016,Rossignoli2019,Rossignoli2022}, we find that the temporal dependence of the cumulative fraction of encounters with Uranus follows a logarithmic function:

\vspace{-3ex}

\begin{equation}
F(t)=a_\text{enc}\ln(t)+b_\text{enc},
\label{ajusfun} 
\end{equation}

\noindent where $a_\text{enc}$\,=\,0.21764$\pm 2 \times10^{-5}$,
$b_\text{enc}$\,=\,-\,3.8311$\pm 4\times10^{-4}$
and $t$ is the time in years ($t>0$, measured from the beginning of the Solar System in its current configuration, see Sect.~\ref{heliomodel}). In addition, a linear function can be fitted to the data for the last 3 Gyr in order to allow comparison with other works on this system (see Sect.~\ref{SecConclusions}). Thus, the linear dependence takes the form: $G(t) = \dot{C}(t-4.5\times10^9)+1$, where $\dot{C}=6.44518\times10^{-11}$ $\pm 3.281\times10^{-14}$ and $t$ is in years. Fig.~\ref{Encur} shows the cumulative fraction of encounters of Centaurs with Uranus over time, together with the logarithmic and linear fits to data.

Considering the logarithmic dependence, the cumulative number of craters for a given satellite as a function of time can be obtained from the following equation:
 \begin{equation}
          N_c(>D,t) = F(t)N_c(>D) ,
          \label{edad2}
 \end{equation}
 
\noindent where $N_c(>D)$ is the total number of craters that have a diameter larger than $D$ produced over the age of the Solar System.

Thus, for the last collision that catastrophically fragmented the satellite at time $t_{\text{rup}}$ we obtain:

\vspace{-3ex}

\begin{equation}
N_c(>D_{\text{rup}},t_f) - N_c(>D_{\text{rup}},t_{\text{rup}})=1 ,
\label{edad3}
\end{equation}    

\noindent where $t_f=$ 4.5 Gyr is the age of the Solar
System and $D_{rup}$ is the crater diameter associated to the catastrophic collision.

In this context, the age of a satellite is defined as the time $\tau_{\text{fr}}$ elapsed since its last catastrophic dispersive collision, or equivalently, the time $\tau_{\text{fr}}$ it has spent in its current orbit without suffering catastrophic dispersive collisions. Thus, using Eq.~\ref{ajusfun}, we obtain:

\begin{equation}
N_c(>D_{\text{rup}})\,\big[a_{\text{enc}}\,(\ln t_f - \ln t_{\text{rup}})\big]=1,
\end{equation}

Consequently, the time interval during which the satellite has avoided catastrophic dispersive impacts in its current orbit is given by:

\begin{equation}
\tau_{\text{fr}} = t_f - t_{\text{rup}} = t_f\Big(1 - e^{-\frac{1}{a_{\text{enc}}N_c(>D_{\text{rup}})}}\Big)
\label{edad11}
\end{equation}

Finally, since our cratering model predicts the cumulative number of impacts over the age of the Solar System, it is necessary to rescale it for satellites that have undergone disruption. For a satellite of age $\tau_{\text{fr}}$, the cumulative number of craters formed since its last catastrophic event is given by:

\begin{equation}
N_c(>D) - N_c(>D,t_{\text{rup}}) = [1 - F(t_{\text{rup}})]N_c(>D)
\label{nuevaprod}
\end{equation}

\subsection{Planetocentric impactor population model}
\label{planetomodel}

The results presented in Sect.~\ref{secresul} suggest that the heliocentric impactors considered in our model may play a secondary role in shaping the mid-sized satellites' observed crater distributions, relative to an alternative, possibly planetocentric impactor population. Based on recent studies that link the current axial tilt of the system and the formation of the regular satellites to a giant impact that occurred in the planet's early history, we follow the study by \citet{Woo2022} to build a planetocentric impactor population. In this scenario, the planetocentric impactors emerge from the final stage of the regular satellites' formation, during which the debris disk from which they accreted was still dynamically excited and contained objects that were potential satellite impactors.

To study this scenario, we followed the general framework of \citet{Woo2022}, using the characteristics of the post-formation debris disk of the mid-sized satellites provided by J.M. Woo (personal communication) as a basis to construct various distributions of particles in planetocentric disks. In their model, the viscous spreading of the post-impact disk generates a positive radial gradient in the solid surface density over the region of satellite formation. In addition, at the end of their $5000\text{ yr}$ N-body simulations, a small population of residual moonlets remains in the outer part of the disk. However, these bodies are few in number and have a resolved size range ($d > 89\text{ km}$) that lies outside the scope of the planetocentric impactor population considered in our model ($d \lesssim 50\text{ km}$, see Eq.~\ref{formaSFD}), preventing us from using their distribution directly. Therefore, we constructed four planetocentric disks to explore a range of profiles with positive gradients. In each disk, the differential number of particles per unit semimajor axis $a$ follows a power law, $dN/da \propto a^n$ where $n = 1, 1.5, 2$, and $2.5$ for disks D1, D1.5, D2, and D2.5, respectively. For each disk, we consider 1000 particles with semimajor axes ranging from 3 R\textsubscript{U} to 30 R\textsubscript{U}.

To account for the excitation of the disk particles after satellite formation, we considered eccentricities in the range $0 \leq e \leq 0.25$ and inclinations between $0^\circ \leq i \leq 15^\circ$ (taking Uranus' north pole as the positive spin axis). The remaining orbital parameters were randomly assigned. We then studied the dynamical evolution of these disks using a numerical simulation performed with the EVORB integrator \citep{Fernandez2002}. The simulations considered massless particles under the gravitational influence of Uranus (including its oblateness) and the five mid-sized satellites in their current orbits. The perturbation exerted by the Sun was also included. The integration time step was $10^{-4}$ years, and the total simulation time was 10 million years. This timespan was chosen based on the rapid dynamical evolution exhibited by these disks. Particles were removed from the simulation upon meeting one of the following conditions: collision with a satellite, reaching a distance $r > 10$ AU from Uranus (ejection from the Uranian system), or approaching the planet within a distance $r < 2.44 \times 10^{-4}$ AU, equivalent to 1.44 $\text{R}_\text{U}$ (Uranus’ Roche radius). Throughout the simulation, the collision parameters between the particles and the mid-sized satellites were recorded.

The results of this simulation, presented in Table \ref{planetour}, provide the collision probabilities for each mid-sized satellite, calculated as the number of collisions each satellite experiences relative to the total number of particles in each disk over the full integration period. Additionally, the recorded collision parameters allowed us to derive the mean impact velocity on each satellite for each disk.

The next step consisted in building an appropriate size-frequency distribution for this impactor population, taking into account that it should reflect a collisionally evolved residual population of planetocentric material remaining after the formation of
the regular satellites. Given that the images of Oberon show terrains dominated by craters, suggesting an ancient surface with a mean crater retention age of 4.5 Gyr with little evidence of endogenic resurfacing \citep{Schenk2020},
we calibrated the SFD of the planetocentric impactor population based on its observed crater density \citep{Kirchoff2022}. First, we converted the observed crater sizes on Oberon into corresponding impactor diameters $d$ by applying the inverse of the crater scaling law (Eqs.~\ref{HH07} and \ref{dcrater}), using the specific mean impact velocity on Oberon obtained for each disk from our EVORB simulations. The resulting impactor SFD from the imaged region \citep{Kirchoff2022} was then scaled to cover Oberon's entire surface area. Finally, to obtain the total number of impactors for each planetocentric disk, the global Oberon impactor SFD was divided by the impact probability on Oberon obtained from the simulation for each disk (see Table \ref{planetour}). This resulted in a different impactor SFD for each disk, given by:
\begin{align}
N(>d) &= a_p \,\bigg(\frac{1 \text{km}}{d}\bigg)^{s_p}
&&\text{for} && d \lesssim 50~\text{km},  
\label{formaSFD}
\end{align} 

\noindent where $d$ is the diameter of the planetocentric impactor and $a_p$ represents the total cumulative number of bodies with $d \ge 1\text{ km}$ in each planetocentric disk. The parameters $a_p$ and $s_p$ differ for each disk, ranging from $a_p=30\,110$ with a cumulative slope $s_p = 1.75$ for disk D1, to $a_p=27\,183$ and $s_p = 1.85$ for disk D2.5. These SFDs enable us to model the crater distribution from planetocentric impactors on the five mid-sized satellites and to compare our results with observational crater counts \citep{Kirchoff2022}. The results of this model are presented in Sect.~\ref{secresul}.

\section{Results}
\label{secresul}
In this section, we present the results of our impact cratering model for the regular Uranian satellites. Initially, we consider only heliocentric impactors for this system, as was done in previous works for the Saturnian satellite system \citep{DiSisto2011,DiSisto2013,Rossignoli2019,Rossignoli2022}. Within this framework, we obtain the crater distributions for all the regular Uranian satellites considering only impacts by the modeled Centaurs and the uncertainty in their SFD, represented by the limiting indices $s_2 = 3$ and $s_2 = 3.5$. Specific results for the case of the largest impactor for both $s_2$ indices are shown in Table \ref{uramedresh} for each regular satellite.  

\begin{table}[h]
\caption{Results for the heliocentric impactor model}
\label{uramedresh}
\begin{flushleft}
\resizebox{\columnwidth}{!}{%
{\begin{tabular}{lcccccc} 
\hline

 \hline\hline\noalign{\smallskip}
Satellite & $v_i$ & $\text{D}_\text{l}$ & $\text{d}_\text{m}$ & $\text{D}_\text{m}$ & $\text{d}_\text{m}$ & $\text{D}_\text{m}$\\
			 & & & {$s_2$=3} & {$s_2$=3} & {$s_2$=3.5} & {$s_2$=3.5}\\
			\hline\noalign{\smallskip}
            Cordelia & 18.95 & 8.83 & 0.27 & 15.88 & 0.88 & 45.68 \\ 
			Ophelia & 18.25 & 8.06 & 0.28 & 16.22 & 0.92 & 46.15 \\ 
            Uranus XXVIII & 17.62 & 37.08 & 0.03 & 1.68 & 0.14 & 8.71 \\
			Bianca & 17.42 & 6.86 & 0.35 & 18.86 & 1.08 & 50.96 \\
			Cressida & 17.06 & 4.52 & 0.52 & 25.55 & 1.50 & 63.07 \\
			Desdemona & 16.94 & 5.30 & 0.44 & 22.57 & 1.32 & 57.66 \\
			Juliet & 16.72 & 3.50 & 0.67 & 30.37 & 1.83 & 71.33 \\
			Portia & 16.52 & 2.65 & 0.88 & 36.68 & 2.27 & 87.85 \\
			Rosalind & 16.08 & 5.15 & 0.45 & 22.14 & 1.32 & 56.53 \\
			Cupid & 15.60 & 20.60 & 0.08 & 5.02 & 0.34 & 19.66 \\
			Belinda & 15.51 & 4.12 & 0.55 & 25.45 & 1.55 & 62.22 \\
			Perdita & 15.40 & 14.26 & 0.16 & 9.15 & 0.57 & 30.45 \\
			Puck & 14.56 & 2.29 & 0.96 & 36.97 & 2.43 & 81.26 \\
			Mab & 13.70 & 30.90 & 0.04 & 2.64 & 0.21 & 12.02 \\
			Miranda & 11.98 & 35.56 & 2.53 & 55.50 & 5.28 & 111.07 \\
			Ariel & 10.04 & 8.43 & 5.78 & 93.33 & 10.22 & 154.96 \\
			Umbriel & 8.66 & 9.93 & 5.41 & 86.95 & 9.71 & 146.33\\
			Titania & 7.02 & 5.67 & 6.69 & 89.41 & 11.49 & 144.29  \\
			Oberon & 6.26 & 6.67 & 6.07 & 81.51 & 10.64 & 134.02\\
		
  \hline
\end{tabular}}
}
\tablefoot{Collision velocity $v_i$ (in kilometers per second), transition crater diameter between the strength and gravity regimes $\text{D}_\text{l}$; largest impactor diameter $\text{d}_\text{m}$ and largest crater diameter $\text{D}_\text{m}$, (all in kilometers) for both values of the SFD index $s_2$=3 and $s_2$=3.5. For the three disrupted inner satellites (Table \ref{fragresults}), the results correspond to the adapted distributions calculated for their specific ages.}
\end{flushleft}
\end{table} 

As mentioned in Sect.~\ref{Secdisrup}, some of the inner satellites have very small sizes, such as Uranus XXVIII, Cupid, and Mab. Thus, following the method described in Sect.~\ref{Secdisrup}, we evaluate whether these bodies experience catastrophic dispersive impacts. Our results show that these three small satellites suffer one or more such events over the age of the Solar System. Consequently, we calculate the time since their last catastrophic dispersive collision (Sect.~\ref{Secedad}), i.e., the age of the satellite. Table~\ref{fragresults} lists our results for these three small satellites, considering the $s_2 = 3.5$ index in the impactors' SFD. In the case of Uranus XXVIII, the parameters were calculated for its largest estimated diameter \citep{ElMoutamid2025} and in the case of Mab, for its smallest estimated diameter \citep[see Sect.~\ref{Pmabsec} and][]{Molter2023}. In addition, we recompute the corresponding crater distributions adapted for the age of each of these satellites using Eq.~\ref{nuevaprod} (see Sect.~\ref{Secdisrup}). Other inner satellites, such as Cordelia, Ophelia, Bianca, and Perdita, may have experienced large impacts that produced craters with diameters comparable to their own size, or that fragmented them without reaching the catastrophic dispersal threshold (see Table ~\ref{uramedresh} and the Appendix).

However, it should be noted that the 14 inner moons may not be primordial nor dynamically stable in their current orbits, as previously discussed \citep{DuncanLi1997,French2012}, although recent work has shown that the Belinda group (Belinda, Perdita, and Cupid) can be stable on 50 Myr timescales via a mean-motion resonance \citep{Cuk2022}. In addition, no craters can currently be identified on these bodies due to the low resolution of the available images, with the exception of Puck (see Fig.~\ref{respuckmab}), preventing a direct comparison between our theoretical predictions and observed crater counts.

In this context, the results presented in this section and in the Appendix, where we show the corresponding calculated crater distributions for all 14 inner satellites, provide constraints on the formation age or the epoch of orbital emplacement of these satellites. These predictions can be tested by future observations, such as those from a dedicated mission to the Uranian system capable of obtaining higher-resolution images of their surfaces. In this regard, the Uranus Orbiter and Probe was identified as the highest-priority new flagship mission in the US Planetary Science and Astrobiology Decadal Survey for the period 2023-2032 \citep{National2023}.

For the mid-sized satellites, the crater distributions produced with the heliocentric impactor model are found to be insufficient to explain the densely cratered surfaces of Umbriel, Titania and Oberon, as shown in Fig.~\ref{UTO}. This suggests the presence of an additional source of impactors within the Uranian system. Thus, we expand the impact cratering model to incorporate the contribution of planetocentric impactors, in addition to the heliocentric population already considered. In this context, we model a set of planetocentric impactor populations linked to the early stages of the dynamical evolution of the system, potentially associated with the debris generated by the giant impact thought to be responsible for the large obliquity of Uranus (see Sect.~\ref{planetomodel}). Such a scenario provides a natural mechanism for producing a transient population of impactors capable of contributing significantly to the observed crater distributions of the mid-sized satellites.

\begin{table}[t]
	\caption{Results for the planetocentric impactor model}
	\begin{flushleft}
     \resizebox{\columnwidth}{!}{%
		{\begin{tabular}{lcccccccc} 
                \hline\hline\noalign{\smallskip}
				Satellite & $v_{\text{mpl}}$ & $p_{\text{col}}$ & $v_{\text{mpl}}$ & $p_{\text{col}}$ & $v_{\text{mpl}}$ & $p_{\text{col}}$ & $v_{\text{mpl}}$ & $p_{\text{col}}$ \\ 
				& D1 & D1 & D1.5 & D1.5 & D2 & D2 & D2.5 & D2.5 \\ 
				\hline\noalign{\smallskip}
				Miranda & 1.51 & 0.020 & 1.62 & 0.010 & 1.72 & 0.007 & 1.94 & 0.003 \\
				Ariel & 1.38 & 0.081 & 1.47 &  0.071 &  1.58 & 0.054 & 1.64 & 0.041 \\
				Umbriel & 1.35 & 0.129 & 1.35 &  0.125 &  1.35 & 0.111 & 1.31 & 0.103 \\
				Titania & 1.19 & 0.341 & 1.20 &  0.325 &  1.20 & 0.363 & 1.20 & 0.353 \\
				Oberon &  1.06 & 0.332 & 1.07 &  0.354 &  1.07 & 0.378 & 1.09 & 0.379 \\
				\hline
		\end{tabular}}
		   }
           \tablefoot{Mean collision velocity $v_{\text{mpl}}$ (in kilometers per second), and collision probability $p_{\text{col}}$ for disks D1 to D2.5 for each mid-sized satellite.}
	\end{flushleft}
	\label{planetour}
\end{table}
To quantify the relative contributions of both populations, we first consider the heliocentric component, modeled through impacts by Centaurs, as was done with the inner satellites (see Table \ref{uramedresh}). We then compute the planetocentric contribution using the disk models described in Sect.~\ref{planetomodel}. Table \ref{planetour} lists the results of the planetocentric model for each disk, presenting the impact probability and impact velocity for each case. Crater distributions obtained with both the heliocentric and planetocentric  models for the mid-sized satellites are presented in Figs. \ref{MirandaAriel} and \ref{UTO}, as cumulative crater densities per square kilometer as a function of crater diameter, together with observed crater distributions. The plots present the heliocentric model results for both $s_2$ values of the impactors' SFD and only the planetocentric model results corresponding to disks D1 and D2.5, as the crater distributions obtained for the intermediate cases (D1.5 and D2) lie between those derived for D1 and D2.5. These two limiting distributions therefore define boundary cases that contain the full range of results predicted by the planetocentric model.

\begin{table}[h]
\caption{\label{t1}Results for catastrophically dispersed satellites}
\label{fragresults}
\centering
\begin{tabular}{lrrr} 
\hline\hline\noalign{\smallskip}
Satellite & $\text{Ds}$ &$d_{rup}$ & $\tau_{fr}$ \\
 & \,[km] & \,[km] & \,[Gyr] \\
\hline
\rule{0pt}{2.6ex}Uranus XXVIII & $\leq$10 & 0.14 & 2.48 \\
            Cupid & 18 & 0.34 & 4.19\\ 
			Mab & 12 & 0.21 & 3.92 \\
\hline
\end{tabular}
\tablefoot{Mean satellite diameter $\text{Ds}$;  diameter of the last disruptive impactor $d_{rup}$, and age of the satellite $\tau_{fr}$.}
\end{table}

In the following, we present a detailed analysis of the results for each mid-sized satellite.

\subsection{Miranda}
On Miranda, \citet{Kirchoff2022} analyzed various regions observed by Voyager 2 using images with resolutions up to 260 meters per pixel, the highest among this group of satellites. The authors identified four distinct regions for crater counting: two cratered areas, ``Cratered Dense'' (40\,180 km$^2$, $5.7\%$ of the satellite's surface) and ``Cratered Low'' (23\,019 km$^2$, $3.3\%$), as well as two coronae: Inverness (41\,266 km$^2$, $5.9\%$  and Elsinore (23\,007 km$^2$, $3.2\%$). For comparison with our modeled crater distributions, we used the ``Cratered Dense'' region (completeness limit CD = 1.3 km, shown in Fig.~\ref{MirandaAriel}) and two of the coronae (completeness limit IC = 1.3 km for Inverness and EC = 1.6 km for Elsinore, shown in Fig.~\ref{MirandaAriel}), excluding "Cratered Low" due to possible image illumination issues or, alternatively, a low crater density possibly linked to recent resurfacing events or impact processes \citep{Kirchoff2022}. Figure \ref{MirandaAriel} shows our results for Miranda using both heliocentric (indices $s_2$=3 and $s_2$=3.5) and planetocentric (disks D1 and D2.5) models. The observed crater distributions from \citet{Kirchoff2022} are also shown for the ``Cratered Dense'' region and the Inverness and Elsinore coronae.

As shown in Figure \ref{MirandaAriel}, the heliocentric model with index $s_2$=3 (solid line) underestimates the number of craters across all sizes in the ``Cratered Dense'' region and most of the coronae, making it inconsistent with observations. In contrast, the model with $s_2$=3.5 (solid line) aligns better with the ``Cratered Dense'' region but does not account for the three largest observed craters on Miranda that have diameters of 44, 51.5, and 86 km. For craters smaller than 10 km, this model overestimates crater counts in the ``Cratered Dense'' region, a discrepancy possibly due to surface renewal processes that more strongly affect smaller craters \citep{Kirchoff2022}. Regarding the coronae, the $s_2$=3.5 model overestimates crater densities across all sizes, consistent with studies suggesting these areas have experienced cryovolcanic resurfacing with elevated heat flow, making them geologically young \citep{Schenk2020}. Notably, the location of the Inverness Corona near Miranda’s south pole resembles that of Enceladus’s most geologically active region, possibly indicating a connection between high heat flow and polar regions \citep{Schenk2020}.

For the planetocentric model, Figure \ref{MirandaAriel} (dashed lines) shows that the D1 disk model matches the three largest craters observed on Miranda, though it overestimates smaller crater counts, likely due to resurfacing effects. Meanwhile, the D2.5 disk model underestimates the crater distribution in the ``Cratered Dense'' region and is therefore considered inconsistent with the observational data.

In summary, a combination of heliocentric impacts with $s_2$=3.5 and planetocentric impacts using a D1 disk model produces a crater distribution consistent with the observations. However, both models tend to overestimate craters smaller than 44 km (planetocentric) or 10 km (heliocentric). Based on this, the crater populations observed in the coronae may be primarily of heliocentric origin, while craters observed in the ``Cratered Dense'' region may have a predominantly planetocentric origin. In fact, this may reflect differences in the ages of these terrains. If the planetocentric impactor population was active only during an early, short-lived phase, its signature would be preserved mainly on older surfaces. In this context, the younger coronae, located in a region affected by resurfacing, would primarily record the cumulative heliocentric collisions over the age of the Solar System, with little contribution from planetocentric impactors. In contrast, the older ``Cratered Dense'' region may retain both components, requiring a combined contribution to reproduce the observed crater distribution, including the largest impact features.

\subsection{Ariel}
\begin{figure}[h]
{\includegraphics[width=0.99\linewidth]{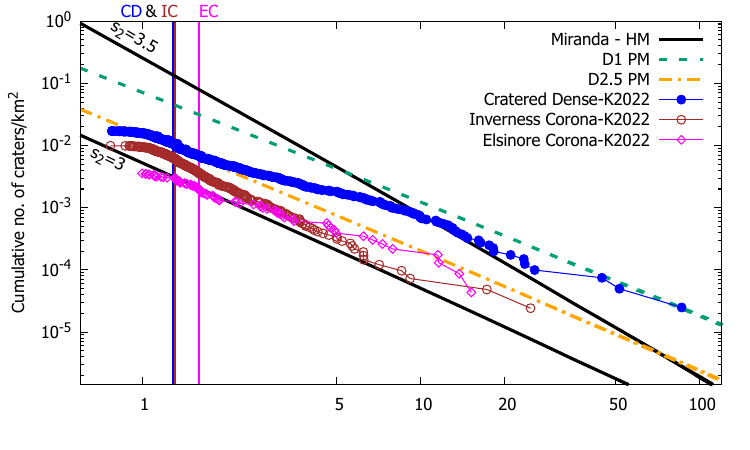}}
{\includegraphics[width=0.99\linewidth]{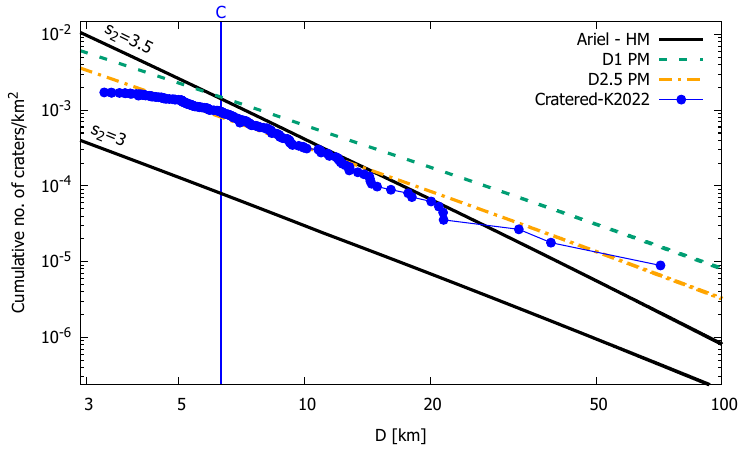}}
\caption{Cumulative number of craters per square kilometer on Miranda (top panel) and Ariel (bottom panel) as a function of crater diameter. Dashed lines represent the crater distributions predicted by our planetocentric model (PM) for disks D1 (D1 PM) and D2.5 (D2.5 PM), black solid lines indicate the crater distributions predicted by our heliocentric model (HM) for both values of $s_2$, and color points indicate observed crater counts \citep{Kirchoff2022}. The vertical colored lines indicate the completeness limit of the crater distributions for each selected region.}
\label{MirandaAriel}
\end{figure}

On Ariel, \citet{Kirchoff2022} studied images with a resolution of 1.3 km/pixel. They identified a cratered region covering 112\,389 km² ($2.6\%$ of the satellite’s surface) and mapped craters with a completeness limit of C = 6.3 km (shown in Fig.~\ref{MirandaAriel}). Overall, the surface appears geologically young, likely shaped by endogenic processes such as cryovolcanism \citep{Schenk2020}. The largest crater, Yangoor (D = 78 km), has a flat topography consistent with degradation processes like viscous relaxation or infill, further supporting the idea of an active surface \citep{Schenk2020}.

Figure \ref{MirandaAriel} presents our results for Ariel using both heliocentric (indices $s_2$=3 and $s_2$=3.5) and planetocentric (disks D1 and D2.5) models, along with the observed crater distribution from the most cratered region reported by \citet{Kirchoff2022}. The heliocentric model with $s_2=3$ (solid line) underestimates crater counts at all sizes and is therefore inconsistent with observations. The $s_2=3.5$ model fits the observed distribution better but does not account for the largest observed crater in the region, Domovoy (D = 71.3 km), and overestimates craters smaller than 10 km, likely due to resurfacing effects that preferentially erase smaller features \citep{Kirchoff2022}.

In the planetocentric case (dashed lines), the D1 disk model overestimates crater numbers across all sizes. Matching this model with observations would require resurfacing processes capable of erasing large craters over $\sim$4.5 Gyr, which may be plausible considering the degraded state of Yangoor. This suggests Ariel may have experienced more large impacts than current imagery reveals. The D2.5 model yields a distribution more consistent with observations but still overestimates crater counts over most of the size range.

In summary, a combination of heliocentric impacts with 
$s_2=3.5$ and planetocentric impacts from disks between D1 and D2.5 yields a crater distribution broadly consistent with observations. However, planetocentric models tend to overpredict crater densities across all sizes. Thus, the heliocentric model appears more consistent with the data, provided Ariel’s surface can retain craters larger than $\sim$8 km over Solar System timescales. Similar to Miranda, this may reflect the timing of the impactor populations and the resurfacing history of Ariel. If the planetocentric population was primarily active during an early, short-lived phase, its contribution would be more difficult to preserve on a surface that has undergone significant endogenic resurfacing. In this context, Ariel’s young and geologically active surface may retain preferentially the impact cratering record of heliocentric origin, while any earlier planetocentric contribution has been partially or largely erased. This is consistent with the better overall agreement of the heliocentric model with the observed crater distribution, particularly when resurfacing processes are taken into account. 

\begin{figure}
{\includegraphics[width=0.99\linewidth]{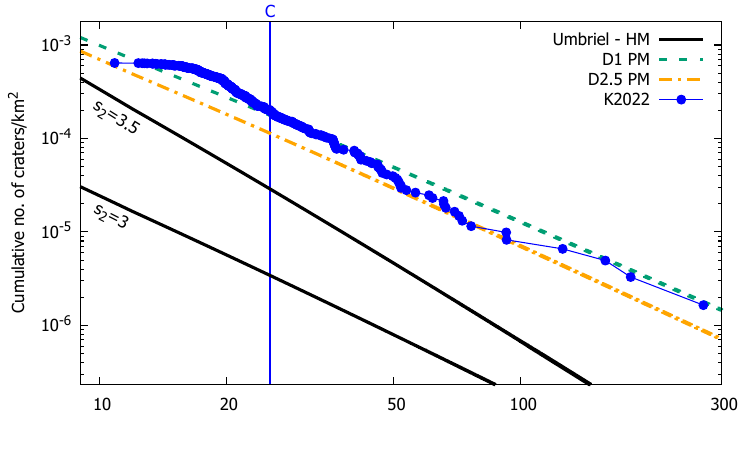}}
{\includegraphics[width=0.99\linewidth]{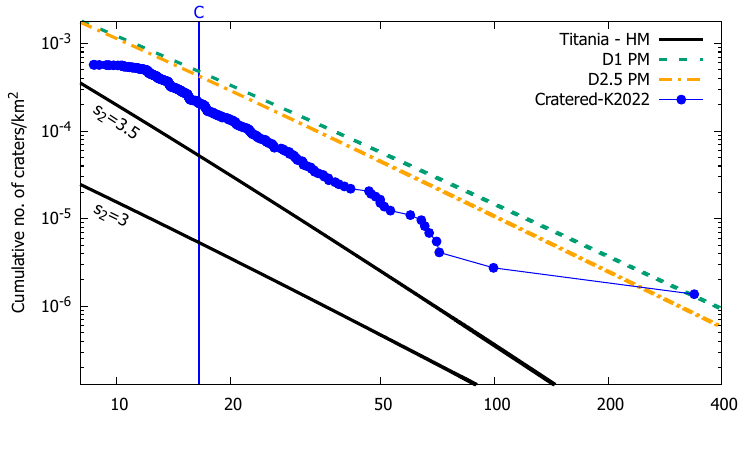}}
{\includegraphics[width=0.99\linewidth]{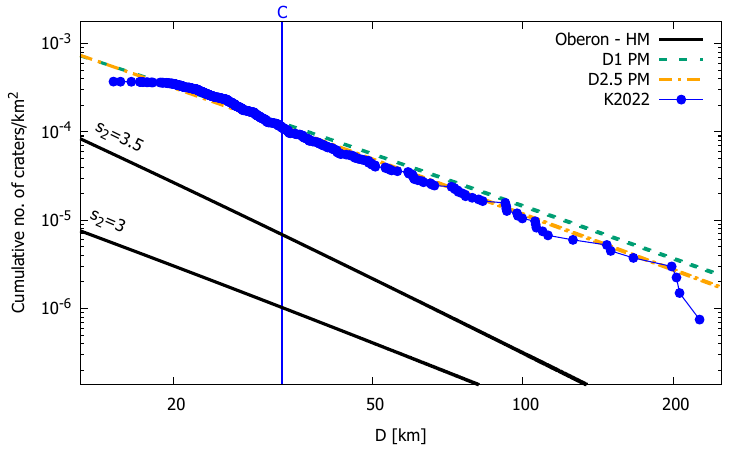}}
\caption{Cumulative number of craters per square kilometer on Umbriel (top panel), Titania (middle panel), and Oberon (bottom panel) as a function of crater diameter. Dashed lines represent the crater distributions predicted by our planetocentric model (PM) for disks D1 (D1 PM) and D2.5 (D2.5 PM), black solid lines indicate the crater distributions predicted by our heliocentric model (HM) for both values of $s_2$, and color points indicate observed crater counts \citep{Kirchoff2022}. The vertical colored lines indicate the completeness limit of the crater distributions for each selected region.}
\label{UTO}
\end{figure}

\subsection{Umbriel}
On Umbriel, \citet{Kirchoff2022} analyzed the surface using Voyager 2 images with a resolution of 5.1 km/pixel. The crater count was performed over an area of 607\,040 km$^2$ ($14\%$ of the satellite's surface), which shows a high crater density consistent with an old, relatively unmodified surface. Due to the image resolution, the crater counts have a completeness limit of C = 25.4 km in diameter (shown in Fig.~\ref{UTO}).

Figure \ref{UTO} presents our results for this satellite obtained with both heliocentric (indices $s_2$=3 and $s_2$=3.5) and planetocentric (disks D1 and D2.5) models, along with the observed crater distribution from \citet{Kirchoff2022}. As shown, the crater distributions obtained with the heliocentric model (solid lines) significantly underestimate the observed crater counts across all sizes, indicating that they are insufficient to explain the observations. For example, at a diameter of $\text{D}=20$ km, the $s_2=3.5$ model predicts 33 craters with $\text{D}\geq20$ km over $14\%$ of the surface, whereas 233 are observed; at 
$\text{D}=50$ km, it predicts 3 compared to 23 observed.

In contrast, the planetocentric model with disk D1 (dashed lines) matches the observed distribution well for craters larger than $\text{D}=20$ km, though it underestimates counts between 10-20 km. Between $40 <\text{D}<125$ km, the model slightly overestimates the number of craters, possibly due to degradation processes such as viscous relaxation affecting large basins \citep{Schenk2020}. Meanwhile, the distribution obtained with the D2.5 model underestimates crater counts across most of the observed size range and is therefore not consistent with the observational data.

In summary, the planetocentric model using disk D1 provides the best fit to the observed crater distribution on Umbriel, despite the underestimation for smaller craters (10–20 km in diameter).

\subsection{Titania}
On Titania, \citet{Kirchoff2022} analyzed Voyager 2 images with a resolution of 3.1 km/pixel. The crater mapping was performed over an area of 726\,903 km$^2$ ($9.3\%$ of the satellite's surface), which shows a lower crater density than Umbriel and Oberon, compatible with a slightly younger terrain. Due to the image resolution, the crater counts have a completeness limit of C = 15.5 km in diameter (shown in Fig.~\ref{UTO}).

Figure \ref{UTO} presents our results for Titania obtained with both heliocentric (indices $s_2$=3 and $s_2$=3.5) and planetocentric (disks D1 and D2.5) models, along with the observed crater distribution from \citet{Kirchoff2022}. As shown, the heliocentric model results for indices $s_2 = 3$ and $s_2 = 3.5$ (solid lines) underestimate the observed number of craters across the entire size range and are therefore insufficient to explain the observations. For example, at $\text{D}=20$ km, the $s_2$=3.5 model predicts 22 craters with $\text{D}\geq20$ km over $9.3\%$ of the surface, while 98 are observed; at $\text{D}=50$ km, the model predicts 2, and 11 are observed.

Regarding the planetocentric model results for disks D1 and D2.5 (dashed lines), both predictions overestimate crater counts at all sizes, except for the largest crater, Gertrude (D = 338.5 km), which aligns with the D1 model prediction.

In conclusion, while both planetocentric models overestimate the observed crater distributions, the D1 disk model provides the closest match to observations on Titania, especially considering that it accurately reproduces the largest crater, Gertrude.

\subsection{Oberon}
On Oberon, \citet{Kirchoff2022} analyzed Voyager 2 images with a resolution of 6.6 km/pixel. The crater mapping was conducted over an area of 1\,336\,217 km$^2$ ($18\%$ of the satellite’s surface), which shows a high crater density consistent with an old surface, minimally altered by endogenic processes. Due to the image resolution, the crater counts have a completeness limit of C = 33 km in diameter (shown in Fig.~\ref{UTO}).

Figure~\ref{UTO} presents our results for both the heliocentric ($s_2 = 3$ and $3.5$) and planetocentric (disks D1 and D2.5) models, compared alongside the observed crater distribution from \citet{Kirchoff2022}. As shown, the heliocentric model results for both $s_2$ indices (solid lines) significantly underestimate the observed number of craters across all diameters and are therefore insufficient to reproduce the observations. For instance, at $\text{D}=20$ km, the $s_2$=3.5 model predicts 35 craters with $\text{D}\geq20$ km over $18\%$ of the surface, while 469 are observed; at $\text{D}=50$ km, the model predicts 3, compared to 57 observed.

In contrast, the planetocentric model for disks D1 and D2.5 (dashed lines) provides distributions consistent with the observations, as the SFDs associated with each disk were calibrated using Oberon's crater record. However, for the three largest observed craters (D > 200 km), both predictions overestimate the observed number of craters.

Overall, the heliocentric model appears to play a clearly secondary role in explaining the cratering on Oberon. For this reason, we developed a planetocentric model calibrated with Oberon’s observed crater distribution and found that, in this system, a planetocentric impactor population like the one modeled could represent the primary source of impacts.

\begin{figure}
		{\includegraphics[width=0.99\linewidth]{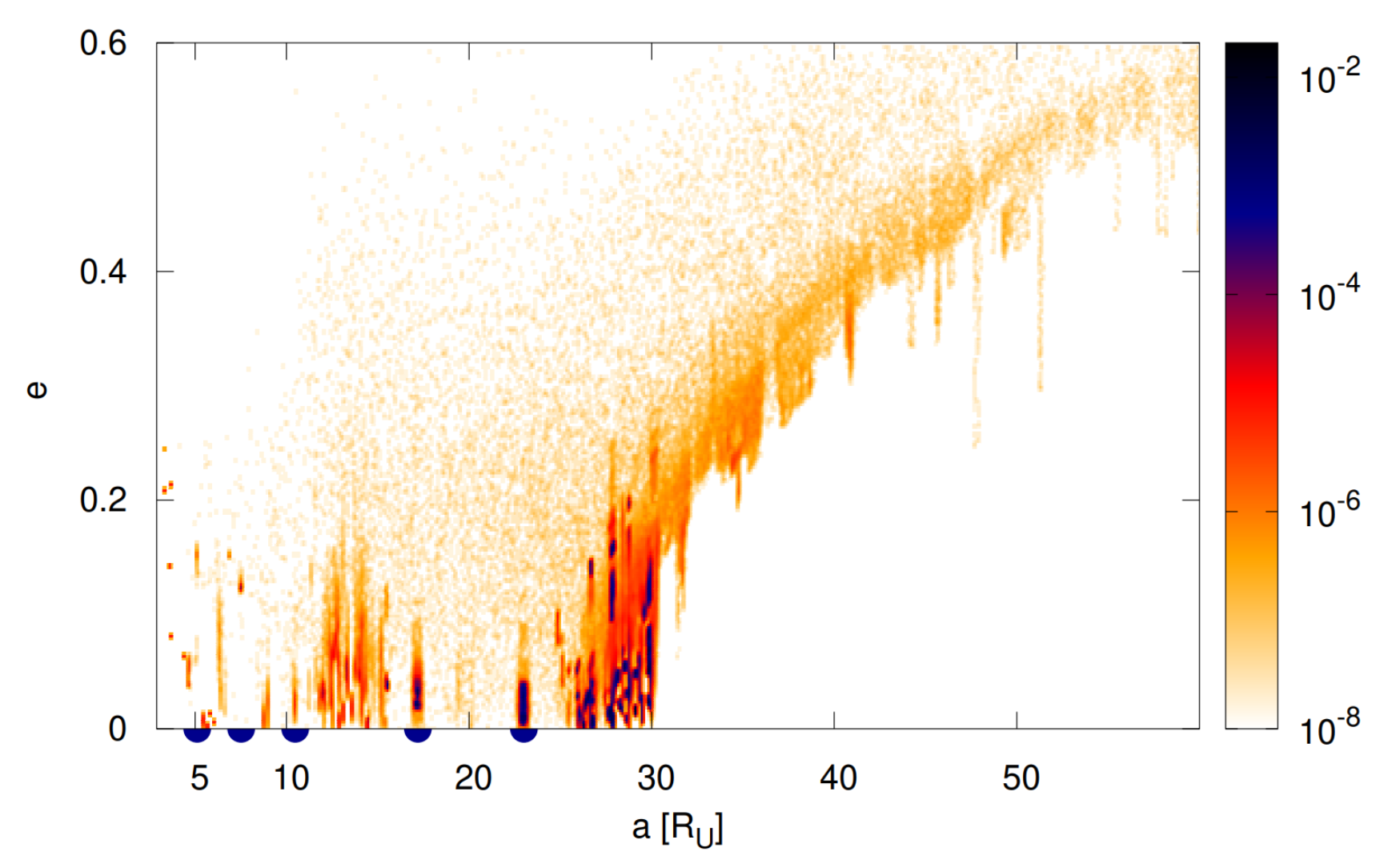}}
			\caption{Time-weighted distribution of the integrated planetocentric impactor population D1 in the semimajor axis (in units of Uranus radii, a ${[\text{R}_{\text{U}}]}$) vs. eccentricity (e) space. Darker colors  denote regions where there is a higher probability of finding impactors throughout the simulation. The locations of the five mid-sized satellites are indicated with dark blue semicircles at the bottom of the plot.}
		\label{map}
	\end{figure}

In addition, from the planetocentric model simulation, a fraction of the simulated particles ($\sim 7\%$) survive after the 10 Myr simulation period. Their orbits lie either within Miranda’s orbit, beyond Oberon’s orbit, or even share orbits with the mid-sized satellites. In Fig. \ref{map} we present the time-weighted distribution of the integrated planetocentric impactor population of Disk D1, in the semimajor axis vs eccentricity space. The color gradient represents the normalized occupation time of each region throughout the 10 Myr simulation, with darker colors indicating the most populated areas. In particular, a number of mean motion resonances are visible, such as the 2:1 and 3:2 with Oberon (at $36.5$ ${\text{R}_{\text{U}}}$ and $30.1$ ${\text{R}_{\text{U}}}$ respectively) and the  2:1 and 3:2 with Titania (at $27.3$ ${\text{R}_{\text{U}}}$ and $22.5$ ${\text{R}_{\text{U}}}$ respectively), as well as particles sharing their orbit with Umbriel, Titania, and Oberon (co-orbitals).  
The survival of long-lived particles could be related to the formation of Uranus' inner satellites and may also point to the existence of still undiscovered moons in the Uranian system. Therefore, a future mission to Uranus is crucial to expand our knowledge on this system, its satellites, and their origin and evolution. 

\section{Discussion and Conclusions}
\label{SecConclusions} 

In the present work, we modeled the impact crater distributions on the regular satellites of Uranus over the age of the Solar System by considering both heliocentric (for all regular satellites) and planetocentric (for the mid-sized satellites) impactor populations. The heliocentric contribution was computed considering Centaurs originating in the SD, using results from numerical simulations of their dynamical evolution \citep{DiSisto2020}. Their SFD was modeled by a power law that takes into account the uncertainties in the small-object population, adopting limiting values for the differential index of $s_2 = 3$ and $s_2 = 3.5$. In addition, we developed a planetocentric impactor model associated with debris generated by a giant impact, with the impactors' SFD calibrated using the observed crater distribution on Oberon. The resulting crater distributions were compared with available observational data for the five mid-sized satellites.

For the inner Uranian satellites, our results obtained with the heliocentric model show that Uranus XXVIII, Cupid, and Mab are expected to undergo catastrophic dispersive collisions over Solar System timescales. This suggests that these objects are unlikely to be primordial in their current form and may instead represent relatively young bodies formed through reaccretion or orbital reconfiguration processes. Future observations of these satellites could provide important constraints on their collisional histories and surface ages.

For the mid-sized satellites, our results show that the crater distributions produced solely with our heliocentric model are not sufficient to reproduce their observed crater densities, particularly for heavily cratered satellites such as Umbriel and Oberon. The heliocentric cratering rates obtained in this work can be compared with previous estimates for the Uranian satellites derived for heliocentric impactors, such as those presented by \citet{Zahnle2003}. Using the linear fit described in Sect.~\ref{Secedad}, we computed cratering rates over the past 3 Gyr and found values broadly consistent with those reported in Table~4 of \citet{Zahnle2003}. For example, adopting $s_2 = 3.5$ in Eq.~\ref{distroCSD}, the cratering rate obtained in this work for Oberon for craters with diameters $D > 10$ km is $1.07\times10^{-14}$ yr$^{-1}$ km$^{-2}$. The corresponding value reported by \citet{Zahnle2003} for the case A impactor population, in which the size-number distribution of impactors was inferred from craters on the Galilean satellites, is $1.4\times10^{-14}$ craters yr$^{-1}$ km$^{-2}$, with an uncertainty factor of six. Similar levels of agreement are obtained for the remaining regular satellites, which supports the consistency of the adopted impactor population and dynamical approach, as was found previously in the Saturnian system \citep[see, e.g.,][]{Rossignoli2022}. Furthermore, our results can be compared with the more recent dynamical model of \citet{Nesvorny2023}, who determined the impact rates by ecliptic comets and Centaurs on the outer planets and their moons. For example, their reported impact rate on Oberon for $d > 1\text{ km}$ impactors is $4.4\times10^{-8}\text{ yr}^{-1}$, while our results considering the linear fit yield a rate of $2.39\times10^{-8}\text{ yr}^{-1}$ for $d > 1\text{ km}$ impactors. This agreement (within a factor of less than two) further supports the consistency of the estimated heliocentric impact rates across independent dynamical models. However, despite this consistency, the heliocentric component of our impact cratering model alone remains insufficient to reproduce the observed crater densities on the mid-sized Uranian satellites. In contrast, the planetocentric impactor populations considered in this work provide a significantly better agreement with the observations, particularly for the most heavily cratered terrains. In the cases of Miranda and Ariel, resurfacing and endogenic activity appear to have modified the crater record substantially in certain regions, reducing or partially erasing the contribution from earlier impactor populations.

These findings suggest that a significant fraction of the cratering record in the Uranian system may be associated with a planetocentric impactor population generated during an early phase of the system’s evolution, while heliocentric impactors provide a continuous background contribution over the age of the Solar System. Recent studies of the bombardment history of the giant planet satellites \citep[e.g.,][]{Wong2019,Wong2021,Bottke2024} have emphasized the importance of early impactor populations and primordial dynamical processes in shaping cratered surfaces throughout the outer Solar System. Within this broader context, our results show that the Uranian system may represent a particularly distinctive case in which debris generated by a giant impact contributed directly to the observed cratering record of the satellites.

Overall, our results suggest that the cratering record of the Uranian satellites cannot be explained solely through the heliocentric impactors considered in our model. Instead, the observed crater distributions may preserve evidence of both heliocentric and planetocentric contributions, together with resurfacing and geological evolution processes that modified the surfaces over time. Future exploration of the Uranian system, including high-resolution imaging and geological characterization of the satellites, will be essential to further refine these conclusions and to better understand the origin and evolution of the Uranian satellite system.

\begin{acknowledgements}
This research was performed using computational resources of Facultad de Ciencias Astronómicas y Geofísicas de La Plata (UNLP) and Instituto de Astrofísica de La Plata (CONICET-UNLP), and received partial funding from Universidad Nacional de La Plata (UNLP) under PID G172. NLR gratefully acknowledges the Asociación Argentina de Astronomía and the Varsavsky family for the Carlos Varsavsky Prize, which contributed to the research presented in this work. We would like to thank Luke Dones for his careful review of the manuscript and for his valuable comments and constructive feedback, which greatly helped improve this paper.     
\end{acknowledgements}

\bibliographystyle{aa}
\bibliography{biblio}

\begin{appendix}
\label{AppenA}
\nolinenumbers
\twocolumn
\section{Results for the inner satellites}

In the following plots we present our calculated crater distributions for all 14 inner satellites. These are shown as cumulative crater numbers per square kilometer as a function of crater diameter for both indices of the heliocentric impactors' SFD. In each case, the lower bound of the curves is determined by the surface area of the satellite, corresponding to a single impact over its entire surface.

\vspace{-2ex}

\subsection*{Cordelia and Ophelia}

\vspace{-4ex}

\begin{figure}[ht!]
\centering
{\includegraphics[width=0.99\linewidth]{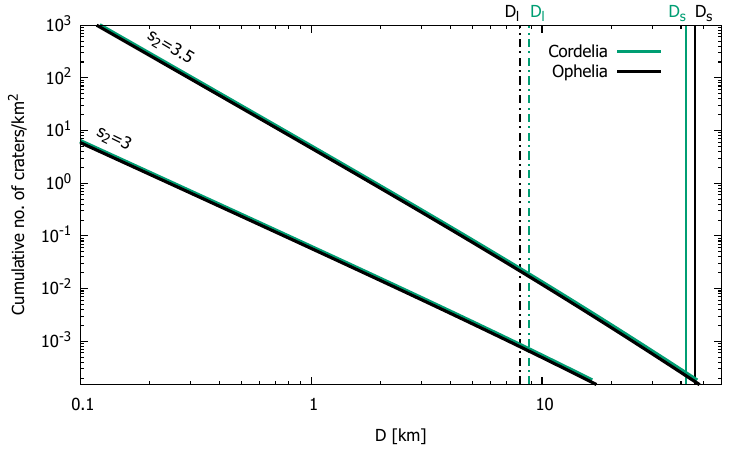}}
\caption{Cumulative number of craters per square kilometer for Cordelia and Ophelia as a function of crater diameter $\text{D}$. Solid lines show the calculated crater distributions for both values of the SFD index. Solid vertical lines indicate each satellite diameter $\text{D}\text{s}$, and dashed lines the transition crater diameter $\text{D}_\text{l}$ from strength to gravity regime.}
\label{rescordof}
\end{figure}

\vspace{-1ex}

Cordelia ($R_{\text{m}}\sim$$21$ km) and Ophelia 
($R_{\text{m}}\sim$$23$ km) are the innermost satellites of Uranus. Their orbital radii differ by only $\sim$4000 km, and together they confine the \(\epsilon\) ring, centered at $\sim$51\,150 km from Uranus. As shown in Fig. \ref{rescordof}, the heliocentric model with $s_2 = 3.5$ yields large impacts on both satellites. The corresponding quantitative results are reported in Table~\ref{uramedresh}. However, in no case does the impact energy exceed the catastrophic dispersal threshold.

\vspace{-2ex}

\subsection*{Uranus XXVIII}
\vspace{-4ex}

\begin{figure}[h!]
\centering
{\includegraphics[width=0.99\linewidth]{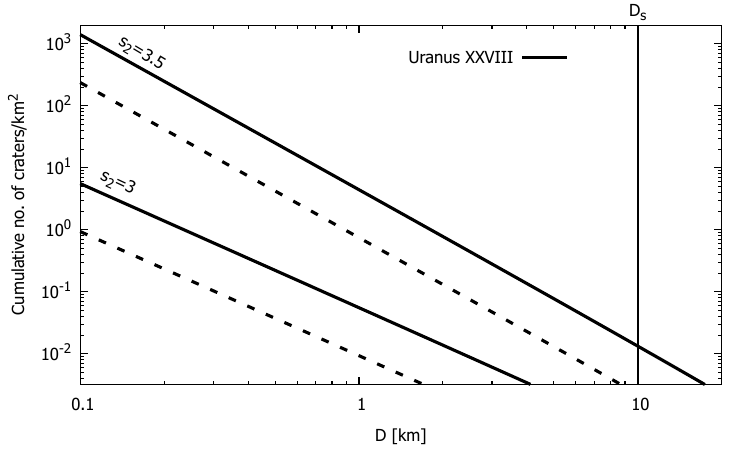}}
\caption{Cumulative number of craters per square kilometer for Uranus XXVIII as a function of crater diameter $\text{D}$. Solid lines show the calculated crater distributions for both values of the SFD index and dashed lines show the distributions scaled to the calculated age of the satellite. The solid vertical line indicates the satellite diameter $\text{D}\text{s}$.}
\label{resuranusxv}
\end{figure}

\vspace{-2ex}
Uranus XXVIII ($R_{\text{m}} \!\sim\! 4-5$ km), a recently discovered inner satellite of Uranus, is the smallest known member of the regular Uranian satellites. Its calculated crater distributions are shown in Fig. \ref{resuranusxv}, which correspond to impacts under the strength regime.

Our results show that this satellite suffers catastrophic heliocentric impacts over the age of the Solar System. According to our model, its most recent catastrophic disruption occurred approximately 2.48 Gyr ago. Its small size implies a short collisional lifetime, consistent with a population of young inner satellites that may be continuously replenished through fragmentation processes. In this sense, objects near the current detection limit may represent the latest generation of small satellites.

\vspace{-2ex}

\subsection*{The Portia group}

\vspace{-3ex}

\begin{figure}[h]
{\includegraphics[width=0.99\linewidth]{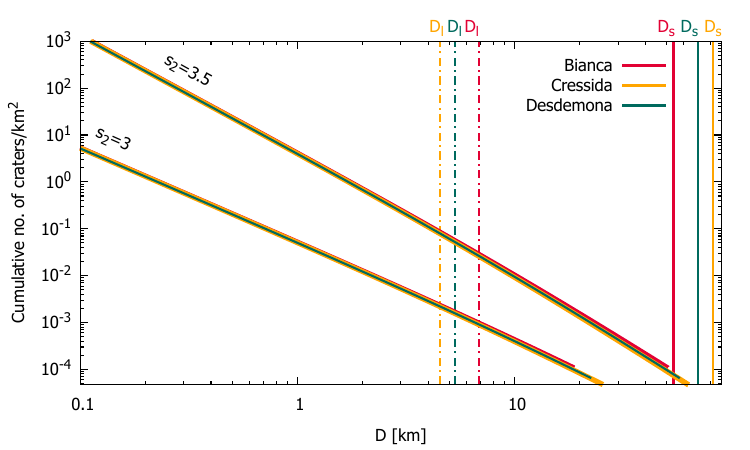}}
{\includegraphics[width=0.99\linewidth]{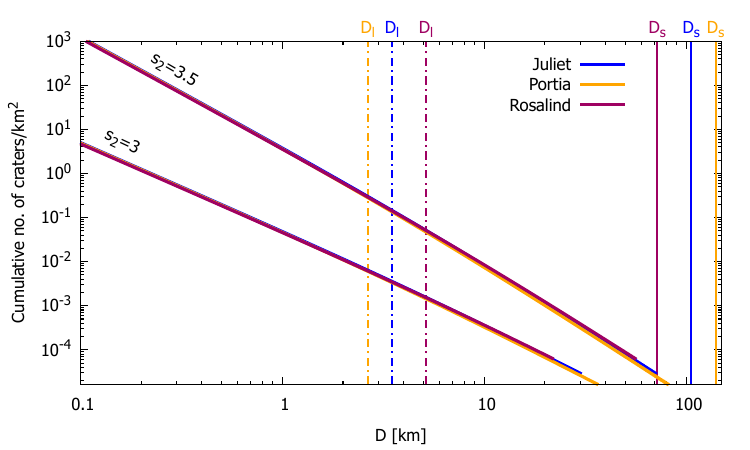}}
{\includegraphics[width=0.99\linewidth]{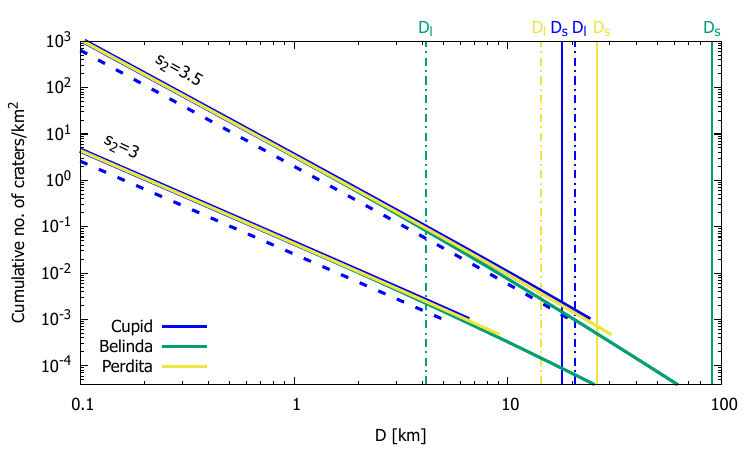}}
\caption{Cumulative number of craters per square kilometer on Bianca, Cressida, and Desdemona (top panel), Juliet, Portia and Rosalind (middle panel), and Cupid, Belinda, and Perdita (bottom panel) as a function of crater diameter. Solid
lines show the calculated crater distributions for both values of
the SFD index, while dashed lines show the distributions scaled to the age of Cupid (bottom panel). Solid vertical lines indicate satellite diameter $\text{D}\text{s}$, and dashed lines the transition crater diameter $\text{D}_\text{l}$ from strength to gravity regime.}
\label{resportia}
\end{figure}

The Portia group comprises nine satellites: Bianca, Cressida, Desdemona, Juliet, Portia, Rosalind, Cupid, Belinda, and Perdita, that present mean radii between $R_{\text{m}} = 9$ km (Cupid) and $R_{\text{m}} = 70$ km (Portia). Figures \ref{resportia} present the results obtained for these satellites, considering impacts produced by the heliocentric model over the age of the Solar System.

In the cases of Bianca (Fig. \ref{resportia}, top panel) and Perdita (Fig. \ref{resportia}, bottom panel), the heliocentric model with index $s_2 = 3.5$ produces large impacts on both satellites, although the catastrophic dispersal threshold is not reached. In contrast, for Cupid (Fig. \ref{resportia}, bottom panel), this limit is exceeded. According to our results, the corresponding age for this satellite is 4.19 Gyr.

\subsection*{Puck and Mab}

\begin{figure}[h!]
{\includegraphics[width=0.95\linewidth]{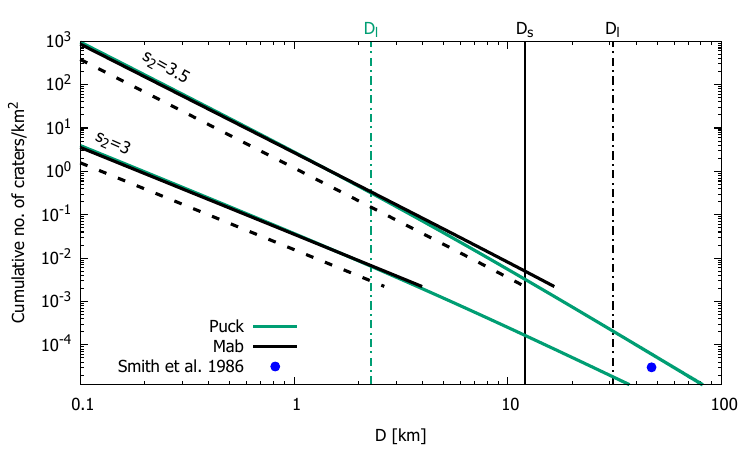}}
\caption{Cumulative number of craters per square kilometer on Puck and Mab as a function of crater diameter. Solid lines show the calculated crater distributions for both values of the SFD index, and dashed lines the distributions scaled to the age of Mab. The solid vertical line indicates the satellite diameter $\text{D}\text{s}$ for Mab, while dashed lines show the transition crater diameter $\text{D}_\text{l}$ from strength to gravity regime. The Bogle crater (47 km diameter) identified by \citet{Smith1986} is shown in blue.}
\label{respuckmab}
\end{figure}

Puck ($R_{\text{m}}=81$ km) and Mab ($R_{\text{m}}=6$ km), associated with the \(\mu\) ring, are the largest and smallest members of this group, respectively. Their results are shown in Fig. \ref{respuckmab}.

For Mab, our results show that it suffers catastrophic dispersive impacts over the age of the Solar System. Its most recent catastrophic disruption occurred 3.92 Ga ago. For Puck, we include the Bogle crater (47 km diameter) identified by \citet{Smith1986}. The $s_2 = 3.5$ model overestimates impacts of this size, which may indicate surface modification processes (e.g., infall of material from the \(\mu\) ring) or that Puck is not primordial, as suggested by \citet{Smith1986}.

\end{appendix}

\end{document}